\documentclass[letterpaper, 10 pt, conference]{ieeeconf}  

\IEEEoverridecommandlockouts                              

\usepackage{cite}
\usepackage{graphicx}
\usepackage{hyperref}
\usepackage{textcomp}
\usepackage{color}
\usepackage{algorithm}
\usepackage{algpseudocode}
\usepackage{amsmath,mathtools,amssymb}
\usepackage{comment}
\usepackage{subcaption}

\newtheorem{theorem}{Theorem}
\newtheorem{lemma}[theorem]{Lemma}
\newtheorem{corollary}[theorem]{Corollary}
\newtheorem{proposition}[theorem]{Proposition}
\newtheorem{remark}{Remark}[section]

\newtheorem{ass}{Assumption}

\newcommand{\R}{\mathbb{R}}

\newcommand\norm[1]{\ensuremath{\lVert#1\rVert}}

\DeclareMathOperator\diag{diag}

\DeclareMathAlphabet\mathbfcal{OMS}{cmsy}{b}{n}

\title{\LARGE \bf
Learning stabilising policies for constrained nonlinear systems
}

\author{Daniele Ravasio, Danilo Saccani, Marcello Farina and Giancarlo Ferrari-Trecate
\thanks{This work was supported as a part of NCCR Automation, a National Centre of Competence in Research, funded by the Swiss National Science Foundation (grant number 51NF40\_225155).}
\thanks{D. Ravasio and M. Farina are with the Dipartimento di Elettronica, Informazione e Bioingegneria (DEIB), Politecnico di Milano, Italy. {\tt\small \{daniele.ravasio, marcello.farina\}@polimi.it}}%
\thanks{D. Ravasio is also with the Istituto di Sistemi e Tecnologie Industriali Intelligenti per il Manifatturiero Avanzato, Consiglio Nazionale delle Ricerche, Italy. {\tt\small daniele.ravasio@stiima.cnr.it}}%
\thanks{D. Saccani and G. Ferrari-Trecate are with the Institute of Mechanical Engineering, Ecole Polytechnique Fédérale de Lausanne (EPFL), CH-1015 Lausanne, Switzerland. {\tt\small \{danilo.saccani, giancarlo.ferraritrecate\}@epfl.ch)}}%
}

\begin{document}

\maketitle
\thispagestyle{empty}
\pagestyle{empty}

\begin{abstract}
This work proposes a two-layered control scheme for constrained nonlinear systems represented by a class of recurrent neural networks and affected by additive disturbances. In particular, a base controller ensures global or regional closed-loop $\ell_p$-stability of the error in tracking a desired equilibrium and the satisfaction of input and output constraints within a robustly positive invariant set. An additional control contribution, derived by combining the internal model control principle with a stable operator, is introduced to improve system performance. This operator, implemented as a stable neural network, can be trained via unconstrained optimisation on a chosen performance metric, without compromising closed-loop equilibrium tracking or constraint satisfaction, even if the optimisation is stopped prematurely. 
Simulations validate the proposed approach on a pH-neutralisation benchmark.
\end{abstract}
%
%
%
%
\section{INTRODUCTION}
Control of constrained nonlinear systems is a long-standing challenge in control theory~\cite{annaswamy2024control}. While stabilising a system is a fundamental requirement, modern applications often demand more: controllers must optimise performance with respect to complex objectives, including economic efficiency, energy consumption, or environmental impact.\\ 
For general nonlinear and nonconvex systems, classical optimal control solutions are scarce and frequently computationally intractable, often relying on restrictive assumptions on the cost function or system structure~\cite{sastry2013nonlinear,pontryagin2018mathematical}. In contrast, optimal control for linear systems with convex cost functions is well understood, but these methods cannot be directly applied to general nonlinear problems.\\
Receding-horizon strategies, such as nonlinear model predictive control (NMPC)~\cite{mayne2000constrained}, provide an effective alternative 
to nonlinear optimal control (NOC) problems. In this context, a finite-horizon open-loop optimal control problem is solved at each sampling instant and the first input is applied to the system, enabling the explicit enforcement of constraints. Despite these advantages, the resulting NMPC law cannot be expressed in closed-form, and many practical applications lack sufficient computational resources to solve nonlinear optimisation problems in real-time.
\\
When experimental data are available, deep neural networks (DNNs) combined with optimal control or reinforcement learning techniques have recently shown great promise in addressing complex control tasks~\cite{mousavi2016deep,berkenkamp2017safe,furieri2025mad,gu2022recurrent}. However, reinforcement learning approaches that guarantee closed-loop stability remain limited and typically rely on restrictive assumptions~\cite{mousavi2016deep,furieri2025mad,berkenkamp2017safe}.  By contrast, constrained NOC relies on the imposition of conservative stability constraints, which can restrict the set of admissible policies or even prevent the synthesis of a viable controller when one exists~\cite{gu2022recurrent}.
\\
A promising alternative is offered by unconstrained optimisation approaches, which exploit system properties to define classes of control policies with built-in stability guarantees~\cite{furieri2022neural,furieri2024learning,galimberti2024parametrizations,saccani2024optimal}. In particular, when the system has $\ell_p$-stable dynamics, the internal model control (IMC) framework can be leveraged to parametrise all and only the stability-preserving control policies. 
This allows one to reformulate the NOC problem as a learning problem by learning over a set of stability-preserving policies parametrised by a stable operator.
Such an operator can be implemented using DNNs with inherent stability guarantees, enabling the problem to be efficiently solved via standard gradient-based techniques without compromising stability, irrespective of the chosen parameter values~\cite{furieri2022neural,furieri2024learning}.\\
If the plant does not inherently possess the desired global stability properties, a suitable base controller must be designed to ensure such properties in closed-loop~\cite{furieri2022neural}. Designing such a controller, however, remains an open challenge for general nonlinear systems. Moreover, global stability properties may not always be admissible due, e.g., to the inherent nonlinearities of the model~\cite{la2024regional}. Finally, certifying constraint satisfaction remains an open challenge with these approaches.

In this work, we address this gap by studying constrained control of a nonlinear plant modelled by a recurrent neural network (RNN) with additive disturbances. RNNs are expressive yet structured models, enabling tractable analysis and controller synthesis~\cite{bonassi2022recurrent}. While prior work has largely focused on certifying stability properties of RNNs~\cite{bonassi2022recurrent,terzi2021learning,revay2023recurrent}, only a few results tackle controller synthesis with formal closed-loop guarantees ~\cite{d2023incremental,ravasio2024lmi,la2024regional,ravasio2026recurrent}.

\subsubsection*{Contributions}
(i) Considering the RNN model class in \cite{la2024regional} and leveraging the results in \cite{ravasio2026recurrent}, we propose a methodology based on linear matrix inequalities (LMIs) to design a pre-stabilising controller that guarantees $\ell_p$-stability of the closed-loop error dynamics and constraint satisfaction within a robustly positive invariant (RPI) set.  
(ii) Inspired by~\cite{furieri2022neural,furieri2024learning,galimberti2024parametrizations}, we propose an IMC-based performance layer, 
which enables optimisation of closed-loop performance while preserving stability even under premature stopping. However, differently from~\cite{furieri2022neural,furieri2024learning,galimberti2024parametrizations}, we exploit the RNN form of the plant model and the features of the prestabilising controller to guarantee constraint satisfaction at all times.
(iii) We characterise the achievable closed-loop responses of the resulting architecture, in which $\ell_p$-stability and constraint satisfaction are guaranteed.
\subsubsection*{Preliminaries}
Given a matrix $M \in \mathbb{R}^{n\times n}\), $m_{ij}\) denotes its $(i,j)\) entry and $M_i$ its $i$-th row. We denote the $n \times n$ identity matrix as $I_n$ and the $n$-dimensional column vector of ones as $1_n$. Let $\R_{\geq 0}$ (resp., $\mathbb Z_{\geq 0}$) be the set of positive real (resp., positive integer) numbers including $0$, $\mathbb R_{+}\coloneqq\mathbb R_{\geq 0}\backslash \{0\}$, and $\mathbb Z_{+}\coloneqq\mathbb Z_{\geq 0}\backslash \{0\}$. Moreover, denote the set of positive definite real matrices as $\mathbb{S}_+^n \coloneqq \{ M \in \mathbb{R}^{n \times n} \ : M=M^\top\succ 0 \}$, and the set of diagonal positive definite real matrices as $\mathbb{D}_+^n$.
Given a square matrix $Q\in\mathbb S_+^n$, we define the set $\mathcal{E}(Q)=\{v\in\R^n:v^\top Qv\leq1\}$.
Given a matrix $H=\diag(h_1,\dots,h_n)$, we define the set $\mathcal I(H)=\{i\in\{1,\dots n\}\,:\,h_i>0\}$.
Given $n$ matrices $M^{(1)},M^{(2)},\dots,M^{(n)}$, we denote by $\diag(M^{(1)},\dots,M^{(n)})$ the block-diagonal matrix having $M^{(1)},\dots,M^{(n)}$ as its main diagonal blocks. 
Given a signal $y(k) \in \mathbb{R}^{n_y}$, with $k\in\mathbb{Z}_{\geq 0}$, $y(k_1:k_2)=(y(k_1),\dots y(k_2))$ for $k_1,k_2 \in \mathbb{Z}_+$ such that $k_1<k_2$. Meanwhile, $\mathbf{y}=(y(0),y(1),\ldots) \in \ell^{n_y}$ denotes the sequence of values taken by $y(k)$ for all $k\geq 0$. Furthermore, the $p$-norm of $\mathbf{y}$ is $\|\mathbf{y}\|_p := (\sum_{j=0}^{\infty} |y(j)|^p)^{1/p}$, for $p\in[1,\infty)$ and  $\|\mathbf{y}\|_\infty := \sup_j |y(j)|$. We say that $\mathbf{y} \in \ell_{p}^{n_y}\subset \ell^{n_y}$ when $\|\mathbf{y}\|_{p}<\infty$.
We use the notation $|\cdot|_p$ to denote the vector $p$-norm and the induced matrix $p$-norm $|A|_p := \sup_{x \neq 0} |A x|_p / |x|_p$. Given the two sequences $\mathbf{y} \in \ell^{n_y}$ and $\mathbf{x} \in \ell^{n_x}$, 
let $\mathbfcal{A}: \mathbf{x} \mapsto \mathbf{y}$ be the operator mapping $\mathbf{x}$ to the $\mathbf{y}$. The operator $\mathbfcal{A}$ is $\ell_p$-stable, i.e., $\mathbfcal{A} \in \mathcal{L}_p$, if it is causal and $\mathbfcal{A}(\mathbf{x}) \in \ell_p^{n_y}$ for all $\mathbf{x} \in \ell_{p}^{n_x}$. Moreover, $\mathbfcal{A} \in \mathcal{L}_p$ has finite $\ell_p$ gain $\gamma(\mathbfcal{A})>0$ if $\|\mathbf{y}\|_p\leq\gamma(\mathbfcal{A}) \|\mathbf{x}\|_p$ holds for all $\mathbf{x} \in \ell_{p}^{n_x}$.  
Finally, given a sequence $\mathbf y$ and a set $\mathcal Y$, we say that $\mathbf y\in\mathcal{Y}^{\mathbb Z_{\geq 0}}$ if $ y(k)\in\mathcal{Y}$ for all $k\in\mathbb Z_{\geq 0}$.
\section{PROBLEM STATEMENT}
\subsection{The plant model}
In this paper we address the problem of controlling a nonlinear plant whose dynamics is described by the following class of RNN models~\cite{la2024regional},
\begin{equation}\label{eq:plant_dynamics}
\begin{cases}
x(k{+}1)\!=\!A_xx(k)\!+\!B_uu(k)\!+\!B_\sigma\sigma(v(k))\!+\!w(k)\\
v(k)=\tilde{A}x(k)+\tilde{B}u(k)\\
y(k)=Cx(k),
\end{cases}\ 
\end{equation}
where $x \in\mathbb{R}^n$ denotes the state vector, $u\in \mathbb{R}^m$ the input vector, $w \in \mathbb{R}^n$ the disturbance vector, and $y\in  \mathbb{R}^{n_y}$ the output vector. 
Matrices $A_x \in \mathbb{R}^{n \times n}$, $B_u \in \mathbb{R}^{n \times m}$, $B_\sigma \in \mathbb{R}^{n \times \nu}$, $\tilde A \in \mathbb{R}^{\nu \times n}$, and $\tilde B \in \mathbb{R}^{\nu \times m}$ are constant model parameters, while 
$\sigma(\cdot) = [\sigma_1(\cdot), \dots, \sigma_\nu(\cdot)]^\top$ is a decentralised vector of sigmoidal functions. We make the following assumptions on system \eqref{eq:plant_dynamics}.
\begin{ass}\label{ass:sigmoid_function}
    Each component $\sigma_i:\R\rightarrow\R$, $i=1,\dots,\nu$, is a sigmoid function, i.e. a bounded, twice continuously differentiable function with positive first derivative at each point and one and only one inflection point in $v_i=0$.
    Also, $\sigma_i(\cdot)$ is Lipschitz continuous with unitary Lipschitz constant and such that $\sigma_i(0)=0$, $\frac{  \partial\sigma_i(v_i)}{\partial v_i}\big|_{v_i=0}=1$ and $\sigma_i(v_i)\in[-1,1]$, $\forall v_i\in\R$.
\end{ass}
\begin{ass}\label{ass:disturbance}
The disturbance sequence $\mathbf{w}$ is drawn from a known distribution $\mathcal{D}$ and satisfies $\mathbf{w} \in \ell_p$. Moreover, each disturbance realisation is uniformly bounded in a known ellipsoid, i.e., $w(k)\in \mathcal{E}(Q_w^0)$, for all $k\geq 0$, where $Q_w^0\in\mathbb S_+^n$.
\end{ass}
For convenience, as in~\cite{la2024regional}, we 
define $A=A_x+B_\sigma \tilde A$, $B=B_u+B_\sigma \tilde B$, $B_q=-B_\sigma$, and the vector $ q(v)=[q_1(v_1),\dots,q_\nu(v_\nu)]^\top$, where $q_i(v_i)=v_i-\sigma_i(v_i)$. The system dynamics \eqref{eq:plant_dynamics} can be reformulated as follows
\begin{equation}\label{eq:error_dynamics1}
    \begin{cases}
        x(k+1)=Ax(k)+Bu(k)+B_qq(v(k))+w(k)\\
        v(k)=\tilde Ax(k)+\tilde Bu(k)\\
        y(k)=Cx(k).
    \end{cases}
\end{equation}
\subsection{The control problem}
Let $(\bar x,\bar u,\bar y)$ be an equilibrium of \eqref{eq:plant_dynamics} such that $\bar x=A_x\bar x+B_u\bar u+B_\sigma \sigma(\tilde A\bar x+\tilde B\bar u)$, $\bar u\in\mathbb U\subseteq \mathbb{R}^{m}$, and $\bar y=C\bar x\in\mathbb Y\subseteq \mathbb{R}^{n_y}$. 
We introduce the following structural assumption on the constraint sets $\mathbb U$ and $\mathbb Y$.
\begin{ass}\label{ass:compact_sets}
The sets $\mathbb U\subseteq\mathbb{R}^m$ and $\mathbb Y\subseteq\mathbb{R}^{n_y}$ are polytopes, i.e., $\mathbb{U}=\{u\in\R^m:G_{\mathrm{u}}u\leq b_{\mathrm{u}}\}$, where $G_{\mathrm{u}}\in\R^{n_t\times m}$ and $b_{\mathrm{u}}\in\R^{n_t}$, and $\mathbb{Y}=\{y\in\R^{n_y}:G_{\mathrm{y}}y\leq b_{\mathrm{y}}\}$, where $G_{\mathrm{y}}\in\R^{n_r\times n_y}$ and $b_{\mathrm{y}}\in\R^{n_r}$.
\end{ass}
The goal of this work is to design a nonlinear, state-feedback, and  possibly time-varying regulator
\begin{equation*}
    \mathbf{u}=\mathbf{R}(\mathbf{x})=(R\big(x(0),0\big),\dots, R\big(x(0{:}k),k\big),\dots)
\end{equation*}
where $\mathbf{R}:\ell^n\rightarrow\ell^m$ is a causal operator that steers the plant state towards a neighbourhood of the equilibrium $\bar x$ (and hence $y(k)=Cx(k)$ towards $\bar y$), minimising a (possibly nonconvex) piece-wise differentiable lower-bounded loss, while satisfying input and output constraints, i.e. $(u,y)\in\mathbb U\times \mathbb Y$, and guaranteeing $\ell_p$ stability of the closed-loop.
More specifically, we aim to solve,
\begin{subequations} \label{eq:Problem1}
\begin{align}
\min_{\mathbf R}\quad &
\mathbb E_{w(0{:}T)}\!\left[\,\mathcal J\big(u(0{:}T),\,y(0{:}T)\big)\,\right] \label{prob:gen-obj}\\
\text{s.t.}\quad &
x(k{+}1)=Ax(k){+}Bu(k){+}B_qq(v(k)){+}w(k)\notag\\
\quad & y(k)=Cx(k), \notag\\
& u(k)= R(x(0{:}k),k)\in\mathbb U,\ \ y(k)\in\mathbb Y, \ \forall k\!\ge\!0, \notag\\
& 
\,\mathbf{y-\bar y}\in\ell_p,\, \mathbf {u-\bar u}\in\ell_p,\label{prob:gen-lp}\\
&\forall\, \mathbf w\in\ell_p\cap\mathcal E({Q_{w}^0})^{\mathbb Z_{\geq 0}}. \nonumber 
\end{align}
\end{subequations}
where $\mathcal J$ denotes a loss function defined over the input and output trajectories $u(0{:}T)$ and $y(0{:}T)$. The expectation $\mathbb{E}_{w(0{:}T)}[\cdot]$ is taken with respect to the disturbance sequence $w(0{:}T)$ and accounts for the fact that performance should be good on average under the disturbance distribution. 
Constraint~\eqref{prob:gen-lp} requires that for any disturbance sequence $\mathbf w$ satisfying Assumption~\ref{ass:disturbance}, the tracking error $\mathbf{y - \bar y}$ and the input deviation $\mathbf{u - \bar u}$ belong to $\ell_p$.\\ 
However, problem \eqref{eq:Problem1} is not directly tractable because the regulator $\mathbf R$ is an infinite-dimensional operator. Moreover, the RNN dynamics together with pointwise input/output constraints lead to a nonconvex optimisation problem.
\subsection{Proposed control scheme}
\begin{figure}[tp]
    \fontsize{8}{12}\selectfont
    \centering
    \def\svgwidth{0.95\columnwidth}
    \input{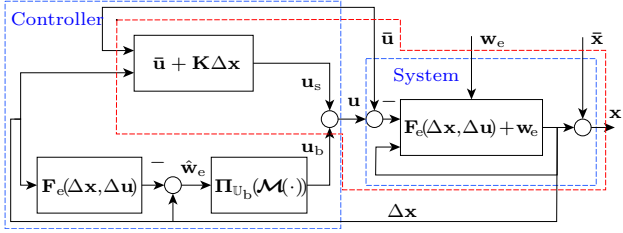}
    \caption{Control architecture. The red dashed block represents the pre-stabilised system.}
    \label{fig:control_architecture}
\end{figure}

To address the control problem~\eqref{eq:Problem1}, we propose a two-layer architecture in which the control input is decomposed as
\[
    u(k)=u_\mathrm s(k)+u_{\mathrm b}(k).
\]
The first component, the stabilising input $u_{\mathrm s}$, is designed for ensuring that the closed-loop system is incrementally input-to-state stable ($\delta$ISS)\footnote{The definitions of $\delta\text{ISS}$ and of dissipation-form $\delta\text{ISS}$ Lyapunov function are provided in~\cite{ravasio2026recurrent}} with respect to both the disturbance $w$ and the additional input channel. The second component, the boosting input $u_{\mathrm b}$, is then used to optimise performance objectives while preserving the stability guarantees enforced by $u_{\mathrm s}$. The overall control scheme is illustrated in Figure~\ref{fig:control_architecture}.
%
%
%
%
%
%
%
%
\section{STABILISING CONTROLLER DESIGN}
The stabilising input $u_\mathrm s$ is defined according to the control law
\begin{equation}\label{eq:control_law_tracking} 
u_{\mathrm{s}}(k)=\bar u+K\Delta x(k),
\end{equation}
where $\Delta x(k)=x(k)-\bar x$, and $K\in\R^{m\times n}$ is the control gain, which is here designed to confer stability properties to the control system.\footnote{In the following, we assume that, at each time instant $k$, an observation $x(k)$ of the system state is available. This assumption is introduced to simplify the framework, although it may be unrealistic in many practical applications. Indeed, since we are dealing with RNN models, the state variables may not correspond directly to measurable physical quantities. To address this issue, the output-feedback procedure proposed in~\cite{ravasio2026recurrent} can be employed to provide an estimate of $x(k)$ and account for the associated uncertainty, without requiring any ad hoc extension of the approach developed in this work.} The closed-loop dynamics given by \eqref{eq:error_dynamics1} under \eqref{eq:control_law_tracking} is described by
\begin{equation}\label{eq:closed_loop_sys}
\begin{cases}
\begin{aligned}x(k+1)=& A_\mathrm{K} x(k)+B(\bar u-K\bar x)+Bu_{\mathrm{b}}(k)\\&+B_q q( v(k))+w(k)\end{aligned}\\
v(k)=\tilde A_\mathrm{K}x(k)+\tilde B(\bar u-K\bar x)+\tilde Bu_{\mathrm{b}}(k)\\
\end{cases}
\end{equation}
where $A_{\mathrm{K}}=A+BK$ and $\tilde A_{\mathrm{K}}=\tilde A+\tilde BK$.\\
To avoid destabilising the inner loop through an overly aggressive performance-boosting action, we restrict the amplitude of the boosting input. Specifically, we assume that $u_\mathrm{b}$ lies in a designer-chosen box $\mathbb U_{\mathrm{b}}\subseteq\mathbb U\ominus \bar u$, defined as
\begin{multline}\label{eq:ub_set}\mathbb{U}_{\mathrm{b}}=\{u_{\mathrm{b}}\in\R^m\,:\,    \begin{bmatrix}G_\mathrm{b}\\-G_\mathrm b\end{bmatrix}u_{\mathrm{b}}\leq1_{2m},\\ G_\mathrm b=\diag(g_{\mathrm{b},1},\dots,g_{\mathrm{b},m})\}.
\end{multline}
where $g_{\mathrm b,i}\in\R_+$, $i = 1,\dots,m$ are design parameters.
Intuitively, $\mathbb{U}_\mathrm{b}$ limits how aggressively the boosting layer may act, and these bounds will later be tuned to balance performance and robustness.\\ 
In the following corollary of \cite[Proposition~4]{ravasio2026recurrent}, the stability properties of \eqref{eq:closed_loop_sys} under a bounded $u_\mathrm{b}$ are stated.
\begin{corollary}\label{cor:dISS}
Let Assumptions \ref{ass:sigmoid_function}, \ref{ass:disturbance}, and \ref{ass:compact_sets} hold. Also, assume that $u_{\mathrm{b}}(k)\in\mathbb U_{\mathrm{b}}$ for $k\geq0$, and define $Q_{w_\mathrm{s}}^0=\diag{(Q_w^0/2,(G_\mathrm{b}^\top G_\mathrm{b})/(2m))}$, $D_{\mathrm s}=[I_n,\,B]$, and $\tilde D_{\mathrm s}=[0,\,\tilde B]$. If there exist $\gamma_\mathrm{s}\in\R_{+}$, and matrices $Q_{\mathrm{s}},\tilde Q_{\mathrm{s,x}},Q_{\mathrm{s},w_\mathrm{s}}\in\mathbb{S}_+^n$, and $H_\mathrm{s},U_\mathrm{s}\in\mathbb D_+^\nu$, with $H_\mathrm{s}=\diag(h_{\mathrm 1},\dots,h_{\mathrm s,\nu})\succeq I_\nu$, and $Z\in\R^{m,n}$ that satisfy the following conditions
\begin{subequations}\label{eq:LMIs}
\begin{equation}\label{eq:dISS_condition}
{\scriptsize
\begin{bmatrix}
    Q_{\mathrm{s}}-\tilde Q_{\mathrm{s,x}}\!\!&\!\!-Q_{\mathrm{s}}\tilde A^\top\!-\!Z^\top \tilde B^\top \!\!&\!\!0\!\!&\!\!Q_\mathrm{s}A^\top\!+\!Z^\top B^\top\\
    -\tilde AQ_\mathrm{s}\!-\!\tilde B Z\!\!&\!\!2H_\mathrm{s}U_\mathrm{s}\!\!&\!\! -\tilde D_{\mathrm{s}}\!\!&\!\!U_\mathrm{s}B_q^\top\\
    0\!\!&\!\!-\tilde D_{\mathrm{s}}^\top \!\!&\!\!Q_{\mathrm{s},w_\mathrm{s}}\!\!&\!\!D_{\mathrm{s}}^\top\\
    AQ_\mathrm{s}\!+\!BZ\!\!&\!\!B_qU_\mathrm{s}\!\!&\!\!D_{\mathrm{s}}\!\!&\!\!Q_{\mathrm{s}}
\end{bmatrix}
    \!\!\succeq 0,}
\end{equation}
\begin{equation}\label{eq:mRPI_set_condition_1_dISS}
{\scriptsize
Q_{\mathrm{s},w_\mathrm{s}}\preceq Q_{w_\mathrm{s}}^0,}
\end{equation}
\begin{equation}\label{eq:mRPI_set_condition_2_dISS}
{\scriptsize
 \tilde Q_{\mathrm{s,x}}-Q_{\mathrm{s}}/\gamma_\mathrm{s}\succeq0,}
\end{equation}
\begin{equation}\label{eq:locality_set_dISS}
{\scriptsize
\begin{bmatrix}\
        Q_{\mathrm{s}}/(2\gamma_\mathrm{s})\!\!&\!\!0\!\!&\!\!Q_{\mathrm{s}}\tilde A_{i}^\top+Z^\top\tilde B_i^\top\\
        0\!\!&\!\!Q_{w_\mathrm{s}}^0/2\!\!&\!\!\tilde D_{\mathrm{s},i}^\top\\
        \tilde A_{i}Q_{\mathrm{s}}+\tilde B_iZ\!\!&\!\!\tilde D_{\mathrm{s},i}\!\!&\!\!(\bar v_i(h_{\mathrm{s},i}){-}|\tilde A_i\bar x{+}\tilde B_i\bar u|)^2
        \end{bmatrix}\!\!\succeq 0, \ \forall i\in\mathcal{I}(H_\mathrm{s}),}
\end{equation}
\begin{equation}\label{eq:out_constr}
{\scriptsize
\begin{bmatrix}\
        Q_{\mathrm{s}}/\gamma_\mathrm{s}\!\!\!&\!\!\!Q_{\mathrm{s}}C^\top G_{\mathrm{y},r}^\top\\
        G_{\mathrm{y},r}CQ_{\mathrm{s}}&\!\!\!(b_{\mathrm{y},r}-G_{\mathrm{y},r}C\bar x)^2
        \end{bmatrix}\!\!\succeq 0, \ \forall r={1,\dots,n_r},}
\end{equation}
\begin{equation}\label{eq:inp_constr}
{\scriptsize
\begin{bmatrix}\
        Q_{\mathrm{s}}/\gamma_\mathrm{s}\!\!&\!\!Z^\top G_{\mathrm{u},t}^\top\\
        G_{\mathrm{u},t}Z\!\!&\!\!(b_{\mathrm{u},t}{-}G_{\mathrm{u},t}\bar u{-}\max_{\tilde u_{\mathrm b\in\mathbb U_{\mathrm{b}}}}G_{\mathrm{u},t}\tilde u_{\mathrm{b}})^2
        \end{bmatrix}\!\!\succeq 0, \ \forall t={1,\dots,n_t},}
\end{equation}
\begin{equation}\label{eq:locality_eq_dISS}
{\scriptsize
\bar v_i(h_{\mathrm{s},i})\geq|\tilde A_i\bar x+\tilde B_i\bar u|, \ \forall i\in\mathcal{I}(H_\mathrm{s}),}
\end{equation}
\end{subequations}
where, for all $i\in\mathcal I(H_{\mathrm s})$,
\begin{equation*}
    \begin{aligned}
    \bar v_i(h_{\mathrm s,i})=&\max_{\tilde v_i}\tilde v_i\\
    \text{s.t }
    &\quad \cfrac{\partial q_i(v_i^\star)}{\partial v_i}\leq\frac{1}{h_{\mathrm s,i}}, \forall v_i^\star\in[-\tilde v_i,\tilde v_i],
    \end{aligned}
    \end{equation*} 
then, setting $K=ZQ_\mathrm{s}^{-1}$ and $P_\mathrm{s}=Q_{\mathrm{s}}^{-1}$, 
\begin{itemize}
    \item if $\mathcal{I}(H_\mathrm{s})=\emptyset$, 
the system described by \eqref{eq:closed_loop_sys} is $\delta$ISS with respect to the sets $\R^n$, $\mathcal E(Q_{w}^0)$, and $\mathbb U_\mathrm b$.
    \item if $\mathcal{I}(H_\mathrm{s})\neq\emptyset$, system \eqref{eq:closed_loop_sys} is $\delta$ISS with respect to the sets $\mathcal{E}(P_\mathrm{s}/\gamma_\mathrm{s})\oplus\bar x$, $\mathcal E(Q_{w}^0)$, and $\mathbb U_\mathrm b$. 
\end{itemize} 
Moreover, the set $\mathcal{E}(P_\mathrm{s}/\gamma_\mathrm{s}) \oplus \bar{x}$ is an RPI set for the closed-loop dynamics, and if $x(0) \in \mathcal{E}(P_\mathrm{s}/\gamma_\mathrm{s})\oplus \bar x$, then $(u(k),y(k)) \in \mathbb{U} \times \mathbb{Y}$ for all $k \geq 0$.
\hfill{}$\square$
\end{corollary}
Corollary~\ref{cor:dISS}, whose proof can be found in the Appendix, provides a systematic procedure for computing $K$ that ensures the $\delta$ISS of the closed-loop system~\eqref{eq:closed_loop_sys}. However, since the associated conditions~\eqref{eq:LMIs} are not LMIs, finding a solution can be challenging. To overcome this difficulty, we propose the following LMI-based procedure:\smallskip\\
\textit{Step 1:} Initialise $G_\mathrm b=\diag(g_{\mathrm b, 1},\dots,g_{\mathrm b,  m})$ such that\begin{align*}
(g_{\mathrm b,1}, \dots, g_{\mathrm b,m})
  &= \arg\max_{g_1, \dots, g_m \in \mathbb{R}_+}
    \sum_{j=1}^m \frac{1}{g_j} \\
&\text{s.t.} \quad
  \sum_{j=1}^m|g_{\mathrm u,i,j}| \, \frac{1}{g_j} \leq \bar b_{\mathrm u,i},
  \quad \forall i = 1, \dots, n_{t}.
\end{align*} 
where $\bar b_\mathrm u = b_\mathrm u - G_\mathrm u \bar u$, and $g_{\mathrm u,i,j}$, 
for $i = 1, \dots, n_t$ and $j = 1, \dots, m$, denotes the $(i,j)$-th entry of $G_\mathrm u$.\\
\textit{Step 2:} Set $H_\mathrm{s} = I_\nu$ and $\gamma_\mathrm{s} = 1$.\\
\textit{Step 3:} Solve the LMI problem~\eqref{eq:dISS_condition}--\eqref{eq:inp_constr}, progressively increasing the values of $h_{\mathrm{s},1},\dots,h_{\mathrm{s},\nu}$ and $\gamma_\mathrm{s}$ until a feasible solution is obtained.\\
\textit{Step 4:} If condition~\eqref{eq:locality_eq_dISS} is satisfied, set $K = Z Q_\mathrm{s}^{-1}$ and $P_\mathrm{s} = Q_\mathrm{s}^{-1}$; otherwise, repeat from Step~2 with increased values of $g_{\mathrm{b},1}, \dots, g_{\mathrm{b},m}$.\smallskip\\ 
The procedure initialises $\mathbb U_\mathrm b$ such that, if $x=\bar x$, then $u=u_\mathrm s+u_\mathrm b=\bar u+u_\mathrm b\in\mathbb U$ for all $u_\mathrm b\in\mathbb U_\mathrm b$.
For the defined set $\mathbb U_\mathrm b$, the procedure initially sets $H_\mathrm{s} = I_\nu$ to enforce closed-loop $\delta$ISS over the largest possible region, and then iteratively updates $H_\mathrm{s}$ and $\gamma_\mathrm s$ to progressively reduce this region until a feasible solution is found. If the region over which $\delta$ISS is enforced is too small for the selected setpoint, the set $\mathbb U_\mathrm b$ is updated and the procedure is repeated. A more detailed discussion on the choice of $H_\mathrm s$ and $\gamma_\mathrm s$ is provided in~\cite{ravasio2026recurrent}. Furthermore, as discussed in~\cite{ravasio2026recurrent}, the existence of a feasible solution to~\eqref{eq:LMIs} is not guaranteed for overly large disturbance sets $\mathcal{E}(Q_w^0)$ or control sets $\mathbb{U}_\mathrm{b}$.\\
Note that the choice of parameters $g_{\mathrm{b},1}, \dots, g_{\mathrm{b},m}$ is particularly critical as it impacts the size of the region of attraction. 
Increasing the set $\mathbb{U}_\mathrm{b}$ generally improves control performance; however, if $\mathbb{U}_\mathrm{b}$ is excessively large, the corresponding RPI set derived from~\eqref{eq:LMIs} may reduce to a size that is impractical for implementation.\smallskip\\
Based on Corollary~\ref{cor:dISS}, if $x(0)\in\mathcal{E}(P_\mathrm s/\gamma_\mathrm s)\oplus \bar x$, the input $u_\mathrm s$ renders the system incrementally stable and guarantees constraints satisfaction, for all $w\in\mathcal E(Q_w^0)$ and $u_\mathrm{b}\in\mathbb U_\mathrm b$.\\
To analyse the equilibrium tracking performance, define $\Delta v = v - \bar v$, $\Delta u = u - \bar u$, $\Delta y = y - \bar y$  and 
$\bar v = \tilde A \bar x + \tilde B \bar u$. 
The resulting closed-loop error dynamics evolves as 
\begin{equation*}
\begin{cases}
\begin{aligned}
&\Delta x(k+1) = A\Delta x(k) + B\Delta u(k) \\
&\qquad + B_q\!\left(q\big(\bar v + \Delta v(k)\big) - q(\bar v)\right) + w(k), \\[4pt]
&\Delta v(k) = \tilde A\Delta x(k) + \tilde B\Delta u(k), \\[4pt]
&\Delta u(k) = K\Delta x(k) + u_{\mathrm b}(k),
\end{aligned}
\end{cases}
\end{equation*}
and, compactly, as
\begin{equation}\label{eq:error_dynamics}
\begin{cases}
\Delta x(k+1) = f_\mathrm{e}(\Delta x(k), \Delta u(k)) + w(k),\\
\Delta u(k) = K\Delta x(k)+u_\mathrm{b}(k).
\end{cases}
\end{equation}
We can now introduce the following proposition for the pre-stabilised system, showing that the $\delta$ISS property of \eqref{eq:closed_loop_sys} implies    $\ell_p$-stability of \eqref{eq:error_dynamics}. 

\begin{proposition}\label{prop:Fcl_lp_plain_noe}
Under Assumption~\ref{ass:disturbance}, consider the closed-loop error dynamics \eqref{eq:error_dynamics}, where $K$ is defined according to Corollary~\ref{cor:dISS}, and define the sequence $\mathbf w_\mathrm e:=(\Delta x(0),w(0),w(1),\dots)$. If 
$\mathbf u_{\mathrm b}\in\ell_p^m\cap\mathbb U_\mathrm b^{\mathbb Z_{\geq 0}}$ and $\Delta x(0)\in\mathcal E(P_\mathrm s/\gamma_\mathrm s)$, then the map
$(\mathbf w_\mathrm e,\mathbf u_{\mathrm b})\mapsto(\Delta\mathbf x,\Delta\mathbf u)$ belongs to $\mathcal L_p$.
\end{proposition}
The proof of Proposition~\ref{prop:Fcl_lp_plain_noe} can be found in the Appendix. In view of it, the closed-loop system~\eqref{eq:closed_loop_sys}
has an $\ell_p$-stable error dynamics for any boosting input
$\mathbf u_{\mathrm b}\in\ell_p\cap\mathbb U_{\mathrm b}^{\mathbb Z_{\geq0}}$.
We are now in a position to design $\mathbf u_{\mathrm b}$ by exploiting the performance-boosting framework proposed in~\cite{furieri2024learning}.
%
%
\section{PERFORMANCE-BOOSTING DESIGN}
The performance-boosting input $u_{\mathrm{b}}$, acting on the system pre-stabilised by $K$, is defined based on an IMC architecture, which comprises an internal model of the closed-loop dynamics, an $\mathcal{L}_p$ operator $\mathbfcal{M}(\cdot)$ designed to optimise the desired performance metric, and a projection operator that guarantees $\mathbf u_{\mathrm{b}} \in \mathbb{U}_\mathrm{b}^{\mathbb Z_{\geq0}}$.\smallskip\\
To formally define the architecture and analyse its properties, we first express the pre-stabilised dynamics \eqref{eq:error_dynamics} in operator form as
\begin{align}
\Delta \mathbf{x} &= \mathbf{F}_{\mathrm{e}}(\Delta \mathbf{x}, \Delta \mathbf{u}) + \mathbf{w}_{\mathrm{e}}, \label{eq:casual_operator_Fk}\\
\Delta \mathbf{u} &= \mathbf{K} \Delta \mathbf{x} + \mathbf{u}_{\mathrm{b}}. \label{eq:casual_operator_policy}
\end{align}
where $\mathbf F_{\mathrm e}:\ell^n\times\ell^m\to\ell^n$ is the strictly causal operator embedding the error dynamics, i.e., $\mathbf F_{\mathrm e}(\Delta\mathbf x,\Delta\mathbf u)=(0,f_\mathrm e(\Delta x(0),\Delta u(0)), f_\mathrm e(\Delta x(1),\Delta u(1)),\dots)$. Moreover, $\mathbf w_\mathrm e= (\Delta x(0),w(0),w(1),\dots)$ represents the sequence of exogenous signals affecting the pre-stabilised closed-loop.\\
We denote a generic causal control policy as $\mathbfcal{K}(\Delta x)$ and define the corresponding closed-loop maps from $\mathbf{w}_\mathrm{e}$ as
\[
\mathbf \Phi^{\Delta\mathbf x}(\mathbf F_{\mathrm e},\mathbfcal K):\ \mathbf w_\mathrm e\mapsto \Delta\mathbf x,
\qquad
\mathbf{\Phi}^{\Delta \mathbf{u}}(\mathbf{F}_{\mathrm{e}}, \mathbfcal{K}) : \mathbf w_\mathrm e \mapsto \Delta\mathbf{u}.
\]
We define the set of achievable closed-loop maps
\begin{equation*}
\mathcal{CL}(\mathbf F_\mathrm e)
= \big\{\,(\boldsymbol{\Phi}^{\Delta\mathbf x}(\mathbf F_{\mathrm e},\mathbfcal K),\boldsymbol{\Phi}^{\Delta\mathbf u}(\mathbf F_{\mathrm e},\mathbfcal K))\big\},
\end{equation*}
and the subset of achievable closed-loop maps that are $\ell_p$-stable and satisfy the constraints as
\begin{multline*}
\mathcal{CL}_p(\mathbf F_\mathrm e,K,\mathcal E(P_\mathrm s/\gamma_\mathrm s),\mathbb U_\mathrm b)
= \big\{\,(\boldsymbol{\Psi}^{\Delta x},\,\boldsymbol{\Psi}^{\Delta u})\in\mathcal{CL}(\mathbf F_\mathrm e)\,:\\\boldsymbol{\Psi}^{\Delta x},\boldsymbol{\Psi}^{\Delta u}\in\mathcal L_p\,\wedge\,
\boldsymbol{\Psi}^{\Delta x}(\mathbf w_\mathrm e)\in\mathcal E(P_\mathrm s/\gamma_\mathrm s)^{\mathbb Z_{\geq 0}}\\ 
\wedge\;
(\mathbf K\boldsymbol{\Psi}^{\Delta x}(\mathbf w_\mathrm e)-\boldsymbol{\Psi}^{\Delta u}\mathbf (\mathbf w_\mathrm e))\in\mathbb U_\mathrm b^{\mathbb Z_{\geq 0}}
\,\big\}.
\end{multline*}
\subsection{The projected internal model control}
The performance-boosting input contribution $\mathbf{u}_{\mathrm{b}}$ is computed as
\begin{equation}\label{eq:pb_control_law}
\mathbf{u}_{\mathrm{b}} = \boldsymbol{\Pi}_{\mathbb{U}_{\mathrm{b}}}\big( \mathbfcal M (\Delta \mathbf x - \mathbf{F}_{\mathrm{e}}(\Delta \mathbf{x}, \Delta \mathbf u))\big),
\end{equation}
where $\boldsymbol{\Pi}_{\mathbb{U}_{\mathrm{b}}}(\mathbf v)$ denotes the projection operator $\boldsymbol{\Pi}_{\mathbb{U}_{\mathrm{b}}} : \ell^m \to \mathbb{U}_{\mathrm{b}}^{\mathbb{Z}_{\ge 0}}$  that acts on a sequence $\mathbf v = (v(0), v(1), \dots)$ by applying a pointwise projection at each time step
, i.e.,
$\Pi_{\mathbb{U}_{\mathrm{b}}}(v(k)) = \arg\min_{\tilde{v} \in \mathbb{U}_{\mathrm{b}}} \| v(k) - \tilde{v} \|_2$ for all $k\in\mathbb Z_{\geq 0}$.
%
%
%
%
\subsection{Main results}
Inspired by~\cite{furieri2022neural,furieri2024learning,galimberti2024parametrizations}, the next theorem formalises the key properties of the proposed performance-boosting architecture. 
\begin{theorem}\label{thm:simple_IMC_local}
Consider the closed-loop error dynamics \eqref{eq:error_dynamics}, where $K$ is defined according to Corollary~\ref{cor:dISS}, and $u_\mathrm{b}$ is defined according to \eqref{eq:pb_control_law}. Under Assumptions~\ref{ass:sigmoid_function}-\ref{ass:compact_sets}, if $\Delta x(0)\in\mathcal E(P_\mathrm s/\gamma_\mathrm s)$, the following statements hold.\smallskip\\
\textbf{(i)} For any $\boldsymbol{\mathcal M}\in\mathcal L_p$ with $\ell_p$-gain $\gamma(\mathbfcal{M})$, the closed-loop maps $\mathbf w_\mathrm e\mapsto \Delta\mathbf x$ and $\mathbf w_\mathrm e\mapsto \Delta\mathbf u$ belong to $\mathcal L_p$. Moreover, the corresponding trajectories satisfy
 $(\mathbf y, \mathbf u) \in (\mathbb{Y} \times \mathbb{U})^{\mathbb{Z}_{\ge 0}}$. 
\smallskip\\
\textbf{(ii)}
For any $(\boldsymbol{\Psi}^{\Delta x},\boldsymbol{\Psi}^{\Delta u})\in\mathcal{CL}_p(\mathbf F_\mathrm e,K,\mathcal E(P_\mathrm s/\gamma_\mathrm s),\mathbb U_\mathrm b)$, there exists $\boldsymbol{\mathcal M}\in\mathcal L_p$ such that \eqref{eq:casual_operator_policy}-\eqref{eq:pb_control_law} achieves these closed-loop maps.
\hfill{}$\square$
\end{theorem}
The proof can be found in the Appendix.
This result shows that, if the system is correctly initialised, then, for any $\boldsymbol{\mathcal{M}} \in \mathcal{L}_p$, the closed-loop system exhibits $\ell_p$-stable error dynamics and satisfies the constraints. 
Also, any causal policy within $\mathcal{CL}_p(\mathbf F_\mathrm e,K,\mathcal E(P_\mathrm s/\gamma_\mathrm s),\mathbb U_\mathrm b)$ can be realised by the same IMC
form with a suitable $\mathbfcal M\in\mathcal L_p$. 
\begin{remark}In the absence of constraints, i.e., $\mathbb{U} \times \mathbb{Y} = \mathbb{R}^m \times \mathbb{R}^p$, if the system \eqref{eq:closed_loop_sys} is $\delta$ISS with respect to $\mathbb{R}^n$, then the set $\mathbb{U}_\mathrm b$ can be chosen arbitrarily large. In this case, the IMC architecture describes all achievable and $\ell_p$-stable maps, i.e. all maps within the set $\mathcal{CL}_p(\mathbf F_\mathrm e,K,\R^n,\R^m)$, thus recovering the global closed-loop results in~\cite{furieri2022neural}.
\hfill$\square$
\end{remark}
%
\subsection{Learning the performance map}
Thanks to Theorem~\ref{thm:simple_IMC_local}, the performance design reduces to optimising $\mathbfcal M \in \mathcal L_p$. 
%
%
Since $\mathcal L_p$ is infinite-dimensional, as in~\cite{furieri2022neural,furieri2024learning,galimberti2024parametrizations}, 
we optimise over a parameterised family $\mathbfcal M(\theta)\!\in\!\mathcal L_p$, where $\theta$ denotes a vector of free parameters. Specifically, we employ a Recurrent Equilibrium Network~\cite{revay2023recurrent}, which guarantees that $\mathbfcal M(\theta)\in\mathcal L_p$ for all $\theta\in\R^{n_\theta}$. Moreover, we replace the expectation in \eqref{prob:gen-obj} with an empirical mean over $S$ disturbance scenarios $\{w^{(s)}\}_{s=1}^S$ drawn from $\mathcal D$. 
The learning problem is formulated as
\begin{align}\label{opt:pb_theta}
\min_{\theta}\quad
& \cfrac{1}{S}\sum_{s=1}^S \,\mathcal J\big(u^{(s)}(0{:}T),\,y^{(s)}(0{:}T)\big)\\
\text{s.t.}\quad
& \Delta x(0)\in\mathcal E(P_{\mathrm s}/\gamma_\mathrm s), \quad w_\mathrm e(0)=\Delta x(0), \notag\\
& w_\mathrm e^{(s)}(k{+}1)=w^{(s)}(k),\notag \\
& \Delta x^{(s)}(k{+}1) {=} f_{\mathrm e}(\Delta x^{(s)}(k),\Delta u^{(s)}(k)) {+} w^{(s)}(k), \notag\\
& \Delta u^{(s)}(k) = K\Delta x^{(s)}(k) + \Pi_{\mathbb{U}_\mathrm b}\!\big(\mathcal{M}(w_\mathrm e^{(s)}(k),\theta)\big), \notag\\
& u^{(s)}(k) = \Delta u^{(s)}(k)+\bar u,\ y^{(s)}(k) = C\Delta x^{(s)}(k)+\bar y \notag\\
& k = 0,\dots,T, \notag
\end{align}
where $\Delta x^{(s)}$ and $\Delta u^{(s)}$ denote the $\Delta x$ and $\Delta u$ trajectories, respectively, when the disturbance $w^{(s)}$ is applied.\\
Note that, due to the
absence of constraints on $\theta$, Problem~\eqref{opt:pb_theta} can be solved efficiently by leveraging standard optimisation frameworks such as PyTorch, using a backpropagation-through-time approach \cite{furieri2022neural,furieri2024learning,galimberti2024parametrizations}. Moreover, since stability is guaranteed by construction and does not require \eqref{opt:pb_theta} to be optimally solved, any piecewise differentiable loss can be used inside $\mathcal{J}$.\smallskip\\
However, the projection operator in \eqref{opt:pb_theta} would, in general, require solving an optimisation problem at each step.
To avoid this, the following lemma provides a closed-form expression of the projection onto the set $\mathbb{U}_\mathrm{b}$ defined in~\eqref{eq:ub_set}.
\begin{lemma} \label{lem:proj}
The projection of $\tilde{u}_\mathrm{b}=\mathcal{M}({w}_\mathrm e)$ onto the set $\mathbb{U}_\mathrm{b}$ 
is given by
\begin{equation}\label{eq:projection} 
\Pi_{\mathbb{U}_\mathrm{b}}(\tilde{u}_\mathrm{b}) 
= G_\mathrm{b}^{-1} \, \mathrm{clip}\!\left(G_\mathrm{b}\tilde{u}_\mathrm{b}\right),
\end{equation}
where, for any vector $v \in \mathbb{R}^n$, the clipping function is defined as
$\mathrm{clip}(v) = \max\!\big(\min(v, 1_n), -1_n\big)$,
with the $\max$ and $\min$ operators applied componentwise. 
\hfill$\square$
\end{lemma}
Lemma~\ref{lem:proj}, whose proof can be found in the Appendix, provides a closed-form expression for the projection, thereby removing the need to solve an optimisation problem at each backpropagation step. Moreover, the resulting operation \eqref{eq:projection} integrates efficiently with gradient-based training frameworks through subgradient computation. 
\section{CASE STUDY}
\begin{figure}[tbp]
     \centering
    \includegraphics[width=0.9\columnwidth]{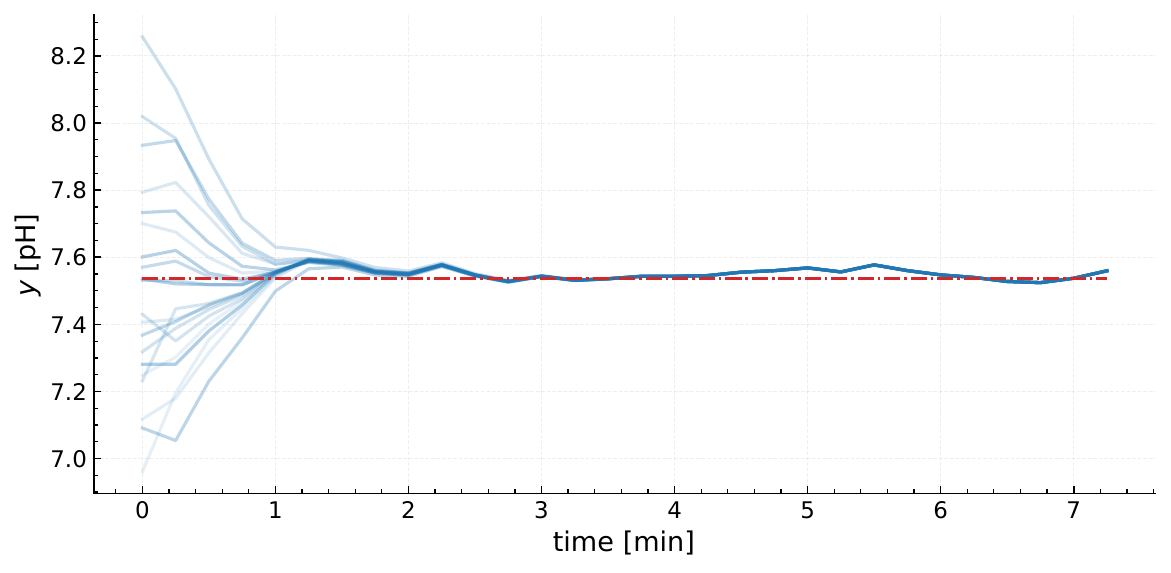}%
    \caption{Closed-loop output performance: system trajectories starting from different initial conditions (blue lines) are compared with the reference (red dash-dotted line).}
    \label{fig:output}
\end{figure}

\begin{figure}[tbp]
     \centering
     \includegraphics[width=0.9\columnwidth]{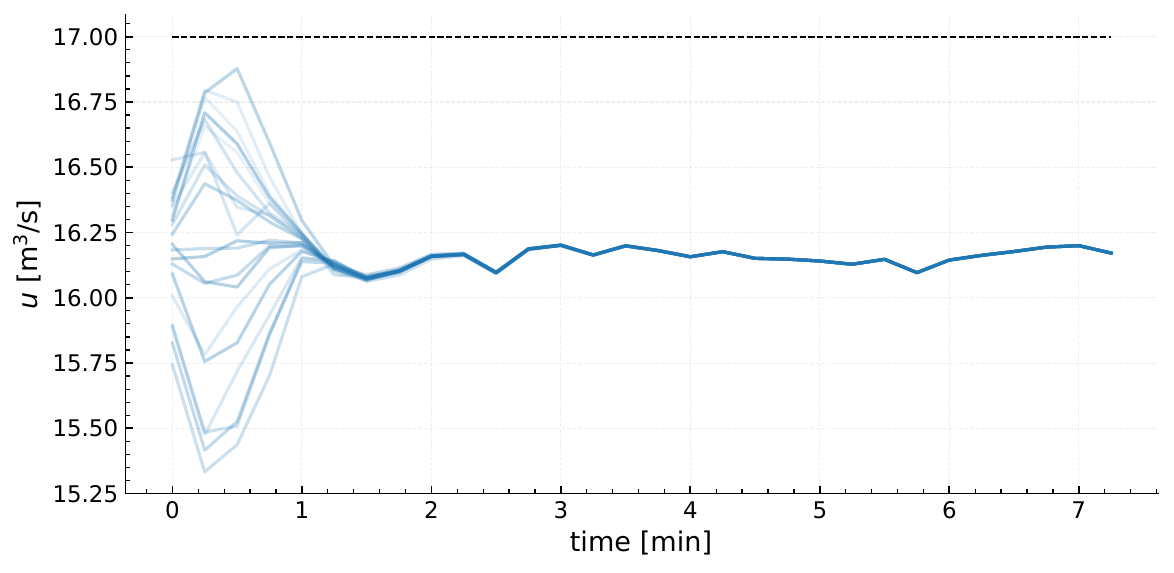}%
    \caption{Evolution of the control input. The black dashed line represents the upper saturation limit.}
    \label{fig:input}
\end{figure}

In this section we evaluate the performance of the proposed approach on the control of a simulation model of the pH-neutralisation process benchmark~\cite{henson2002adaptive}.\\
The pH-neutralisation process is a nonlinear single-input single-output system. 
The controllable input $u$ is the alkaline base flow rate, while the measured output $y$ is the pH of the output flow rate. 
Saturation constraints are considered on both the input and the output, i.e., $u \in [12.5,\ 17]$ and $y \in [5.94,\ 9.13]$. 
The equations of the model, alongside its parameters, are reported in~\cite{henson2002adaptive}.\\
First, an input–output dataset with a sampling period of $15$~s has been obtained by exciting the simulator with a multilevel pseudo-random signal designed to cover different operating regions and frequencies. 
An RNN-based model of the class \eqref{eq:plant_dynamics}, with $n = 10$ states and activation functions $\sigma_i = \tanh(\cdot)$ for $i = 1, \dots, 5$, has been identified from the normalised dataset, and used to design the proposed control algorithm.  To test the robustness of the approach, simulations have been carried out in the presence of a bounded disturbance $\mathbf w\in\ell_\infty$ such that $\|\mathbf{w}\|_\infty<0.01$. 
The set for the performance-boosting input has been selected as $\mathbb{U}_\mathrm{b,\mathrm{norm}}=[-u_M, u_M]$ in normalised units, where $u_M=0.0912$. Also, we consider the following loss function: 
\begin{multline*}
\mathcal{J}
= \sum_{k=0}^T\sum_{s=0}^S\Big(\omega_1 \big| 10^{-y^{(s)}(k)} - 10^{-\bar y} \big|
\\+ \omega_2 \big| u^{(s)}(k) - u^{(s)}(k-1) \big| 
+ \omega_3 \, \max(|\tilde u^{(s)}_\mathrm{b}(k)-u_\mathrm{M}|,0 )\Big)
\end{multline*}
where $\omega_1=1/10^{-\bar y}$, $\omega_2=0.1$, and $\omega_3=0.05$. The first term in the loss function penalises deviations of the hydrogen ion concentration $10^{-y^{(s)}}$ from the target concentration $10^{-\bar y}$. Note that expressing the tracking error in terms of concentration accounts for the logarithmic pH scale and emphasises acidic deviations, which have a higher environmental impact.  The second term in the loss function discourages large changes in the inlet alkaline flow. Finally, the last term penalises violations of the safety bound for $u_\mathrm b$.\\
Note that, due to the first term in the loss, $\mathcal J$ is nonconvex. This term is in general approximated by a quadratic tracking term, penalising the pH tracking error $y-\bar y$, see e.g. \cite{ravasio2026recurrent}, neglecting the environmental impact of acid deviations to make the optimisation problem tractable.\smallskip\\
Figures~\ref{fig:output}--\ref{fig:input} display the closed-loop simulation results starting from different initial conditions but subject to the same disturbance realisation, previously unseen during the performance boosting controller training. 
In particular, Figure~\ref{fig:output} shows that output effectively tracks the setpoint $\bar{y}$. Moreover, the figure 
highlights the effect of the performance-boosting: before reaching the setpoint, all trajectories are pushed above the reference, i.e., towards more alkaline pH values. Comparison with the case in which the input $u_\mathrm{b}$ is optimised using a standard tracking error index shows a mean reduction of the acid deviations, averaged over 20 trajectories, of $7.9$\%. 
Finally, Figure~\ref{fig:input} shows that the control input evolution always remains within the prescribed constraints.
%
%
%

\section*{CONCLUSIONS}
In this work, we proposed a control architecture for constrained nonlinear systems modelled by a class of RNNs. The framework enables the unconstrained optimisation of control performance while guaranteeing robust closed-loop stability and constraint satisfaction.
A base controller is designed to ensure constraint satisfaction and $\ell_p$-stability of the state-tracking error within an RPI set. Moreover, an additional control contribution, derived from the IMC principle, is optimised over a freely chosen loss function and projected onto a designer-specified set to safely improve the closed-loop performance.
Future work will focus on extending the proposed scheme to enlarge the controller’s region of attraction. 

\bibliographystyle{IEEEtran}
\bibliography{biblio}

\section{Appendix: Proof of the main results} %
In this appendix we report the proofs of the main results reported in the paper.\smallskip\\
\textbf{Proof of Corollary~\ref{cor:dISS}.}
The first part of the proof relies on \cite[Proposition~4]{ravasio2026recurrent}.\\
Defining $w_\mathrm{s} = [w^\top,\, u_\mathrm{b}^\top]^\top$, the dynamics \eqref{eq:closed_loop_sys} can be rewritten as $x^+= A_\mathrm{K} x+B(\bar u-K\bar x)+D_\mathrm s w_\mathrm s+B_q q(\tilde A_\mathrm{K} x+\tilde B(\bar u-K\bar x)+\tilde D_\mathrm s w_\mathrm s)$, that is equivalent to (10) in~\cite{ravasio2026recurrent}.\\ 
Since $u_{\mathrm{b}} \in \mathbb{U}_\mathrm{b}$ if and only if $g_{\mathrm{b},i} |u_{\mathrm b,i}| \leq 1$ for all $i = 1, \dots, m$, it follows that  
\[u_{\mathrm{b}}^\top G_\mathrm{b}^\top G_\mathrm{b} u_{\mathrm{b}} = \sum_{i=1}^m (g_{\mathrm{b},i} u_{\mathrm b,i})^2 \leq m,\]
that is, $u_{\mathrm{b}} \in \mathcal{E}(G_\mathrm{b}^\top G_\mathrm{b}/m)$.
Therefore, under Assumption~\ref{ass:disturbance}, it follows that 
$w^\top Q_w^0 w + u_{\mathrm{b}}^\top ((G_\mathrm{b}^\top G_\mathrm{b})/m) u_{\mathrm{b}} \leq 2$, 
which is equivalent to 
$w_\mathrm{s} \in \mathcal{E}(Q_{w_\mathrm{s}}^0)$.\\  Suppose that conditions~\eqref{eq:dISS_condition}--\eqref{eq:locality_set_dISS} and~\eqref{eq:locality_eq_dISS} hold. Then, the $\delta$ISS properties of \eqref{eq:closed_loop_sys}, as well as the invariance of $\mathcal{E}(P_\mathrm{s}/\gamma_\mathrm{s}) \oplus \bar{x}$, follow from \cite[Proposition~6]{ravasio2026recurrent}.\smallskip\\
We now show that if \eqref{eq:out_constr} holds and $x \in \mathcal{E}(P_\mathrm{s}/\gamma_\mathrm{s})$, then $y \in \mathbb{Y}$. Left- and right-multiplying \eqref{eq:out_constr} by $\mathrm{diag}(P_\mathrm{s},1)$, we obtain
\begin{equation}\label{eq:out_constr_P}
\begin{bmatrix}\
        P_{\mathrm{s}}/\gamma_\mathrm{s}\!\!\!&\!\!\!C^\top G_{\mathrm{y},r}^\top\\
        G_{\mathrm{y},r}C&\!\!\!(b_{\mathrm{y},r}-G_{\mathrm{y},r}C\bar x)^2
        \end{bmatrix}\!\!\succeq 0, \ \forall r\in\{1,\dots,n_r\},
\end{equation}
According to \cite[Lemma~10]{ravasio2026recurrent}, condition \eqref{eq:out_constr_P} implies that $|G_{\mathrm{y},r} C \Delta x| \leq b_{\mathrm{y},r} - G_{\mathrm{y},r} \bar{x}$ for all $r = 1, \dots, n_r$. This further implies $G_{\mathrm{y},r} C (x - \bar{x}) \leq b_{\mathrm{y},r} - G_{\mathrm{y},r} \bar{x}$ for all $r = 1, \dots, n_r$. Rearranging the latter inequality yields $G_{\mathrm{y},r} C x \leq b_{\mathrm{y},r}$ for all $r = 1, \dots, n_r$, i.e., $y \in \mathbb{Y}$.\smallskip\\
Finally, we show that if \eqref{eq:inp_constr} holds, $x \in \mathcal{E}(P_\mathrm{s}/\gamma_\mathrm{s})$, and $u_{\mathrm{b}} \in \mathbb{U}_\mathrm{b}$, then $u \in \mathbb{U}$. Left- and right-multiplying \eqref{eq:inp_constr} by $\mathrm{diag}(P_\mathrm{s},1)$ and recalling that $Z = KQ$, we obtain, for all $t\in\{1,\dots,n_t\}$,
\begin{equation}\label{eq:inp_constr_P}
\begin{bmatrix}\
        P_{\mathrm{s}}/\gamma_\mathrm{s}&K^\top G_{\mathrm{u},t}^\top\\
        G_{\mathrm{u},t}K&(b_{\mathrm{u},t}-G_{\mathrm{u},t}\bar u-\max_{\tilde u_{\mathrm b\in\mathbb U_{\mathrm{b}}}}G_{\mathrm{u},t}\tilde u_{\mathrm{b}})^2
        \end{bmatrix}\!\!\succeq 0.
\end{equation}
Then, according to \cite[Lemma~10]{ravasio2026recurrent}, condition \eqref{eq:inp_constr_P} implies that
$|G_{\mathrm{u},t} K(x - \bar{x})| \leq b_{\mathrm{u},t} - G_{\mathrm{u},t} \bar{u} - \max_{\tilde u_{\mathrm{b}} \in \mathbb{U}_\mathrm{b}} G_{\mathrm{u},t} \tilde u_{\mathrm{b}}$, for all $t = 1, \dots, n_t$.
Noting that $\max_{\tilde u_{\mathrm{b}} \in \mathbb{U}_\mathrm{b}} G_{\mathrm{u},t} \tilde u_{\mathrm{b}} \geq G_{\mathrm{u},t} u_{\mathrm{b}}$, this further implies, for all $t = 1, \dots, n_t$, 
\begin{align*}
G_{\mathrm{u},t} K(x - \bar{x}) &\leq b_{\mathrm{u},t} - G_{\mathrm{u},t} \bar{u} - \max_{\tilde u_{\mathrm{b}} \in \mathbb{U}_\mathrm{b}} G_{\mathrm{u},t} \tilde u_{\mathrm{b}} \\
&\leq b_{\mathrm{u},t} - G_{\mathrm{u},t} \bar{u} - G_{\mathrm{u},t} u_{\mathrm{b}}.
\end{align*} 
Rearranging the latter inequality and substituting \eqref{eq:control_law_tracking} yields 
$G_{\mathrm{u},t} (u_\mathrm{s} + u_{\mathrm{b}}) \leq b_{\mathrm{u},t}$, for all $t = 1, \dots, n_t$, i.e., $u = u_\mathrm{s} + u_{\mathrm{b}} \in \mathbb{U}$.
\hfill{}$\square$\smallskip\\
\textbf{Proof of Proposition~\ref{prop:Fcl_lp_plain_noe}.}
From Corollary~\ref{cor:dISS}, given the trajectories $(x,w,u_{\mathrm{b}}),(\bar x,0,0)\in(\mathcal E(P_\mathrm s/\gamma_\mathrm s)\oplus \bar x)\times\mathcal E(Q_w^0)\times\mathbb U_{\mathrm{b}}$, it holds $V(\Delta x^+)-V(\Delta x)\leq-\sigma_x\norm{\Delta x}^2+\sigma_{w_\mathrm s}(\norm{w}^2+\norm{u_{\mathrm{b}}}^2)$, where $V(\Delta x)=\Delta x^\top P_\mathrm s\Delta x$, $\sigma_x=\lambda_{\mathrm{min}}(P_\mathrm s\tilde Q_{\mathrm{s,x}}P_\mathrm s)$ and $\sigma_{w_\mathrm s}=\lambda_{\mathrm{max}}(Q_{w_\mathrm s})$. It follows that
$V(\Delta x^+)-V(\Delta x)\leq -(\sigma_x/\lambda_{\mathrm{max}}(P_\mathrm s)) V(\Delta x)+\sigma_{w_\mathrm s}(\| w\|^2+\| u_{\mathrm{b}}\|^2)$, that is $V(\Delta x^+)\leq \tilde a V(\Delta x)+\sigma_{w_\mathrm s}(\| w\|^2+\| u_{\mathrm{b}}\|^2)$, where $\tilde a=1-\sigma_x/\lambda_{\mathrm{max}}(P_\mathrm s)\in[0,1)$.\\
By iteration, we obtain
\begin{multline*}
V(\Delta x(k))\ \le\ \tilde a^{k}V(\Delta x(0))
\\+ \sum_{j=0}^{k-1} \tilde a^{k-j-1}\sigma_{w_\mathrm{s}}\left(\|w(j)\|^2 + \|u_{\mathrm{b}}(j)\|^2\right).
\end{multline*}
Since $\lambda_{\mathrm{min}}(P_\mathrm s)\norm{\Delta x}^2\leq V(\Delta x)\leq\lambda_{\mathrm{max}}(P_\mathrm s)\norm{\Delta x}^2$, it follows that
\begin{multline*}
\norm{\Delta x(k)}^2\ \le\ \cfrac{\lambda_{\max}(P_{\mathrm s})}{\lambda_{\min}(P_{\mathrm s})}\tilde a^{k}\norm{\Delta x(0)}^2
\\+ \cfrac{\sigma_{w_\mathrm{s}}}{\lambda_{\min}(P_{\mathrm s})}\sum_{j=0}^{k-1} \tilde a^{k-j-1}\left(\|w(j)\|^2 + \|u_{\mathrm{b}}(j)\|^2\right),
\end{multline*}
that implies \begin{align*}
\norm{\Delta x(k)}&\leq \sqrt{\cfrac{\lambda_{\max}(P_{\mathrm s})}{\lambda_{\min}(P_{\mathrm s})}\tilde a^{k}\norm{\Delta x(0)}^2}
\\&\quad+ \sqrt{\cfrac{\sigma_{w_\mathrm{s}}}{\lambda_{\min}(P_{\mathrm s})}\sum_{j=0}^{k-1} \tilde a^{k-j-1}\left(\|w(j)\|^2 + \|u_{\mathrm{b}}(j)\|^2\right)}\\
&\leq \kappa_0 a^{k}\norm{\Delta x(0))}
\\&\quad+ \kappa_1\sum_{j=0}^{k-1} a^{k-j-1}\left(\|w(j)\| + \|u_{\mathrm{b}}(j)\|\right),
\end{align*}
where $a=\sqrt{\tilde a}\in[0,1)$, $\kappa_0^2=\lambda_{\max}(P_{\mathrm s})/\lambda_{\min}(P_{\mathrm s})$ and $\kappa_1^2=\sigma_{w_\mathrm s}/\lambda_{\mathrm{min}} (P_\mathrm s)$.\\
Taking the $p$-norm for $p\in[1,\infty)$ and using Minkowski’s inequality,
\begin{align*}
\|\Delta \mathbf x\|_p
&\le \Big\|\big(\kappa_0 a^{k}\|\Delta x(0)\|\big)_{k\ge 0}\Big\|_p
   \\
   & \qquad \quad+ \Big\|\big(\kappa_1\sum_{j=0}^{k-1} a^{k-1-j}(\|w(j)\|+\|u_{\mathrm b}(j)\|)\big)_{k\ge 0}\Big\|_p \\
&= \kappa_0\,\mu_p\,\|\Delta x(0)\| \;+\; \kappa_1\,\|\rho * \eta\|_p,
\end{align*}
where the symbol ``\(*\)'' denotes the discrete-time convolution operator, i.e.,
$(\rho * \eta)_k = \sum_{j=0}^k \rho_j \eta_{k-j}$ with $\rho_j:=a^{j}$, $\eta_j:=\|w(j)\|+\|u_{\mathrm b}(j)\|$ and $\mu_p = \| (a^k )_{k\geq0} \|_p$. By Young’s inequality,
$\|\rho*\eta\|_{p}\le \|\rho\|_{1}\|\eta\|_{p}$ and $\|\rho\|_{1}=\sum_{j\ge 0}a^{j}=1/(1-a)$. Hence
    $\|\Delta\mathbf x\|_{p}\le \kappa_0\,\mu_p\,\|\Delta x(0)\|
+\frac{\kappa_1}{1-a}\big(\|\mathbf w\|_{p}+\|\mathbf u_{\mathrm b}\|_{p}\big).$
Recalling the stacked exogenous definition $\mathbf w_\mathrm e=(\Delta x(0),w(0),w(1),\ldots)$, for any $p\in[1,\infty]$ we have 
$\|\Delta x(0)\|\ \le\ \|\mathbf w_\mathrm e\|_p$ and 
$\|\mathbf w\|_p\ \le\ \|\mathbf w_\mathrm e\|_p$, 
because for $p<\infty$ the $p$-norm of the concatenation satisfies
$\|\mathbf w_\mathrm e\|_p^p=\|\Delta x(0)\|^p+\|\mathbf w\|_p^p$
and for $p=\infty$ it is the supremum over all components.
Substituting into the previous bound yields
\begin{equation}
\label{eq:bound_deltax}
    \|\Delta\mathbf x\|_{p}\ \le\ \big(\kappa_0\,\mu_p+\tfrac{\kappa_1}{1-a}\big)\,\|\mathbf w_\mathrm e\|_{p}
\;+\; \tfrac{\kappa_1}{1-a}\,\|\mathbf u_{\mathrm b}\|_{p}.
\end{equation}
Taking the $p$-norm to $\Delta \mathbf u$ in~\eqref{eq:error_dynamics}, we have
\begin{align}
\|\Delta\mathbf u\|_p\ &\le\ |K|_p\,\|\Delta\mathbf x\|_p+\|\mathbf u_{\mathrm b}\|_p,\notag \\
&\le |K|_p\big(\kappa_0\,\mu_p+\tfrac{\kappa_1}{1-a}\big)\,\|\mathbf w_\mathrm e\|_{p}\label{eq:inputboundLp}\\
&\quad
+ (|K|_p\tfrac{\kappa_1}{1-a}+1)\,\|\mathbf u_{\mathrm b}\|_{p}\notag.
\end{align}
The case $p=\infty$ follows identically with $\mu_\infty=1$. \\Since $(\mathbf w_\mathrm e,\mathbf u_\mathrm b)\in \ell_p^n \times \ell_p^m$ by assumption, inequalities \eqref{eq:bound_deltax}–\eqref{eq:inputboundLp} imply that $(\Delta \mathbf x,\Delta \mathbf u)\in \ell_p^n \times \ell_p^m$. 
As the error dynamics in \eqref{eq:error_dynamics} is strictly causal, the induced map $(\mathbf w_\mathrm e,\mathbf u_\mathrm b) \mapsto (\Delta \mathbf x,\Delta \mathbf u)$ belongs to $\mathcal{L}_p$.
\hfill$\square$\smallskip\\
\textbf{Proof of Theorem \ref{thm:simple_IMC_local}.}
\emph{(i)} 
Because projection onto a nonempty closed convex set in a Hilbert space is $1$-Lipschitz~\cite{boyd2004convex},
and since the projection is applied componentwise over time (and $0\in\mathbb U_{\mathrm{b}}$), we have $\|\mathbf u_{\mathrm{b}}\|_{p}=\|\boldsymbol{\Pi}_\mathbb{U_\mathrm b}( \mathbfcal M(\mathbf w_\mathrm e))\|_{p}\leq\|\mathbfcal M(\mathbf w_\mathrm e)\|_{p}\le \gamma(\mathbfcal M)\|\mathbf w_\mathrm e\|_{p}$.
Then, the first claim of (i) follows from Proposition~\ref{prop:Fcl_lp_plain_noe}.
Moreover, \eqref{eq:pb_control_law} implies that $ u_{\mathrm{b}}\in\mathbb U_{\mathrm b}^{\mathbb Z_{\geq0}}$. Therefore, the last claim of \emph{(i)} follows from Corollary~\ref{cor:dISS}.\smallskip\\
\emph{(ii)} Let $(\boldsymbol{\Phi}^{\Delta x}, \boldsymbol{\Phi}^{\Delta u})$ be the closed-loop maps generated by \eqref{eq:casual_operator_Fk}, \eqref{eq:casual_operator_policy}, and \eqref{eq:pb_control_law}, where we define the operator $\mathbfcal M (\cdot)$ as
$\mathbfcal M (\cdot)= -\mathbf K \boldsymbol{\Psi}^{\Delta \mathbf x}(\cdot)+\boldsymbol{\Psi}^{\Delta \mathbf u}(\cdot)$. Since $(\boldsymbol{\Psi}^{\Delta \mathbf{x}}, \boldsymbol{\Psi}^{\Delta \mathbf{u}}) \in \mathcal{CL}_p$, it holds that $\boldsymbol{\Psi}^{\Delta \mathbf{x}},\boldsymbol{\Psi}^{\Delta \mathbf{u}}\in\mathcal{L}_p$ and  $(\mathbf{K}\boldsymbol{\Psi}^{\Delta \mathbf{x}}(\mathbf w_\mathrm e) - \boldsymbol{\Psi}^{\Delta \mathbf{u}}(\mathbf w_\mathrm e)) \in \mathbb{U}_\mathrm{b}^{\mathbb{Z}_{\geq 0}}$. Therefore, $\boldsymbol{\mathcal{M}} \in \mathcal{L}_p$. Moreover, since $\mathbb{U}_\mathrm{b}$ is defined as a symmetric box around the origin, it follows that $\mathbfcal{M}(\mathbf{w}_\mathrm{e}) \in \mathbb{U}_\mathrm{b}^{\mathbb{Z}_{\geq 0}}$. Using \eqref{eq:pb_control_law}, this implies that 
\begin{equation}\label{eq:pb_law_induction}
\mathbf{u}_\mathrm{b} \!=\! \boldsymbol{\Pi}_{\mathbb{U}_\mathrm{b}}(\mathbfcal{M}(\mathbf{w}_\mathrm{e})) \!=\! -\mathbf{K}\boldsymbol{\Psi}^{\Delta \mathbf{x}}(\mathbf{w}_\mathrm{e}) + \boldsymbol{\Psi}^{\Delta \mathbf{u}}(\mathbf{w}_\mathrm{e})
\end{equation}
For $k=0$, we have $\Phi^{\Delta x}(w_\mathrm e(0)) = \Delta x(0)=\Psi^{\Delta x}(w_\mathrm e(0))$. Then, \eqref{eq:casual_operator_policy} and \eqref{eq:pb_law_induction} imply $\Phi^{\Delta u}(w_\mathrm e(0)) = K \Phi^{\Delta x}(w_\mathrm e(0)) + u_b(0) = K \Phi^{\Delta x}(w_\mathrm e(0)) - K\Psi^{\Delta x}(w_\mathrm e(0)) + \Psi^{\Delta u}(w_\mathrm e(0)) = \Psi^{\Delta u}(w_\mathrm e(0))$. A simple induction argument on $k$ (as in the proof of~\cite[Thm.~2]{furieri2022neural}) concludes that the two sequences coincide for all $k\geq0$.
\hfill{}$\square$\smallskip\\
\textbf{Proof of Lemma \ref{lem:proj}}:
For convenience, we rewrite the set $\mathbb{U}_\mathrm{b}$ in~\eqref{eq:ub_set} as a zonotopic set with generator matrix 
$G_\mathrm{z} = G_\mathrm{b}^{-1}=\mathrm{diag}(1/g_{\mathrm{b},1}, \dots, 1/g_{\mathrm{b},m})$, i.e.,
\[
\mathbb{U}_\mathrm{b} = \{ u_{\mathrm{b}} = G_\mathrm{z} \, \xi \,:\, \| \xi \|_\infty \leq 1, \, \xi\in \R^m \}.
\]
Using the zonotopic representation of $\mathbb{U}_\mathrm{b}$, the projection of $\tilde{u}_\mathrm{b}$ can be computed as 
$\Pi_{\mathbb{U}_\mathrm{b}}(\tilde{u}_\mathrm{b}) = G_\mathrm{z} \xi^\star$, where
\[
\xi^\star = \arg\min_{\xi \in [-1_m, 1_m]} \| \tilde{u}_\mathrm{b} - G_\mathrm{z} \xi \|.
\]
Let $\xi^\mathrm{u}$ denote the unconstrained minimiser, i.e.,
\[
\xi^\mathrm{u} = \arg\min_{\xi\in\R^m} \| \tilde{u}_\mathrm{b} - G_\mathrm{z} \xi \|.
\]
Since $G_\mathrm{z}$ is diagonal, with diagonal entries $g_{\mathrm{z},i} = 1 / g_{\mathrm{b},i}$ for $i = 1, \dots, m$, 
the optimisation problem decouples across components, i.e.,
\[
\xi^\mathrm{u}_i = \arg\min_{\xi_i} \, g_{\mathrm{z},i}^2 \xi_i^2 - 2 g_{\mathrm{z},i} \tilde{u}_{\mathrm{b},i} \xi_i 
= \frac{\tilde{u}_{\mathrm{b},i}}{g_{\mathrm{z},i}}, \quad i = 1, \dots, m.
\]
Projecting $\xi^\mathrm{u}$ onto the the set $[-1_m, 1_m]$ yields
$\xi^\star = \mathrm{clip}(\xi^\mathrm{u})$, 
and substituting back gives
\[
\Pi_{\mathbb{U}_\mathrm{b}}(\tilde{u}_\mathrm{b}) 
= G_\mathrm{z} \, \mathrm{clip}\!\left(G_\mathrm{z}^{-1} \tilde{u}_\mathrm{b}\right),
\]
which is equivalent to \eqref{eq:projection} recalling that $G_\mathrm b=G_\mathrm z^{-1}$.
\hfill{}$\square$\smallskip\\

\end{document}